\documentclass{article}
\usepackage{spconf,amsmath,graphicx}

\usepackage{enumitem}
\setlist{nosep, leftmargin=14pt}

\usepackage{mwe} 

\usepackage{booktabs}
\usepackage{url}
\usepackage{setspace}
\usepackage{multirow}
\usepackage[pagebackref,breaklinks,hidelinks]{hyperref}
\usepackage[capitalize]{cleveref}

\title{Sequential Spatial-Temporal Network for Interpretable Automatic Ultrasonic Assessment of Fetal Head during labor}
\name{\begin{tabular}{c}Jie Gan$^{1}$, Zhuonan Liang$^{1}$, Jianan Fan$^{1}$, Lisa Mcguire$^{2}$,\\
  Caterina Watson$^{3}$, Jacqueline Spurway$^{4}$, Jillian Clarke$^{2}$, Weidong Cai$^{1,*}$\end{tabular}}
  \address{$^{1}$ School of Computer Science, The University of Sydney, NSW, Australia\\
           $^{2}$ Sydney School of Health Sciences, The University of Sydney, NSW, Australia\\
           $^{3}$ School of Science, Edith Cowan University, WA, Australia\\
           $^{4}$ Medical Imaging Department, Orange Hospital, NSW, Australia}
\begin{document}
%
\maketitle
\begin{abstract}
  The intrapartum ultrasound guideline established by ISUOG highlights the Angle of Progression (AoP) and Head Symphysis Distance (HSD) as pivotal metrics for assessing fetal head descent and predicting delivery outcomes. Accurate measurement of the AoP and HSD requires a structured process. This begins with identifying standardized ultrasound planes, followed by the detection of specific anatomical landmarks within the regions of the pubic symphysis and fetal head that correlate with the delivery parameters AoP and HSD. Finally, these measurements are derived based on the identified anatomical landmarks. Addressing the clinical demands and standard operation process outlined in the ISUOG guideline, we introduce the Sequential Spatial-Temporal Network (SSTN), the first interpretable model specifically designed for the video of intrapartum ultrasound analysis. The SSTN operates by first identifying ultrasound planes, then segmenting anatomical structures such as the pubic symphysis and fetal head, and finally detecting key landmarks for precise measurement of HSD and AoP. Furthermore, the cohesive framework leverages task-related information to improve accuracy and reliability. Experimental evaluations on clinical datasets demonstrate that SSTN significantly surpasses existing models, reducing the mean absolute error by 18\% for AoP and 22\% for HSD.
\end{abstract}
\begin{keywords}
  Temporal processing, Sequential Tasks, labor, Intrapartum Ultrasound Video, AoP and HSD
\end{keywords}
\section{Introduction}
\label{sec:intro}
\begin{figure}[ht]
  \centering
  \includegraphics[width=0.43\textwidth]{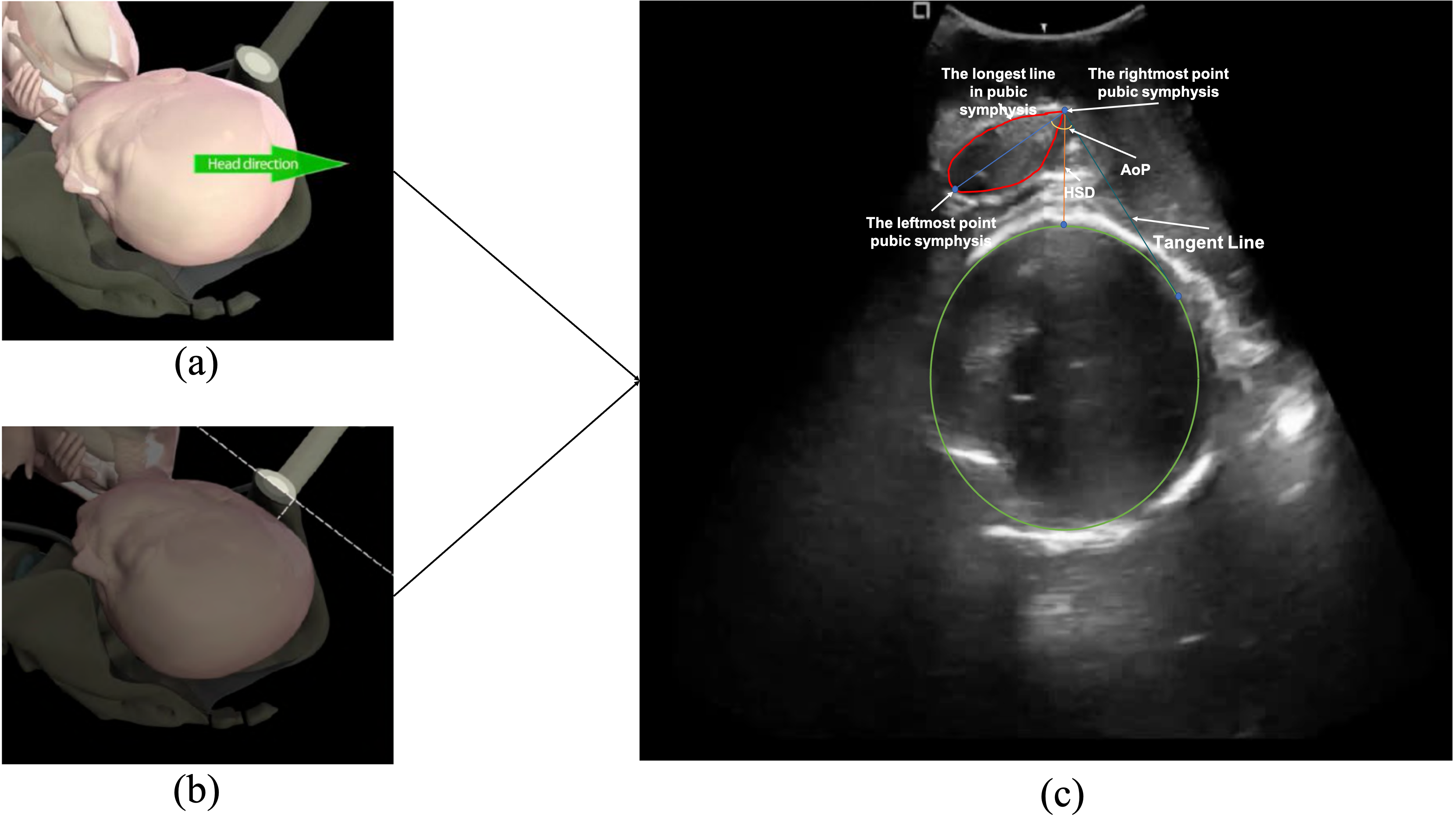}
  \caption{The illustration of fetal head assessment by AoP and HSD. Free-head positioning to measure the AoP (a) and HSD (b). The ultrasound biometric (c) is measured by following the protocols in (a) and (b).}
  \label{fig:aop&hsd}
\end{figure}

\begin{figure*}[ht]
  \centering
  \includegraphics[width=0.65\textwidth]{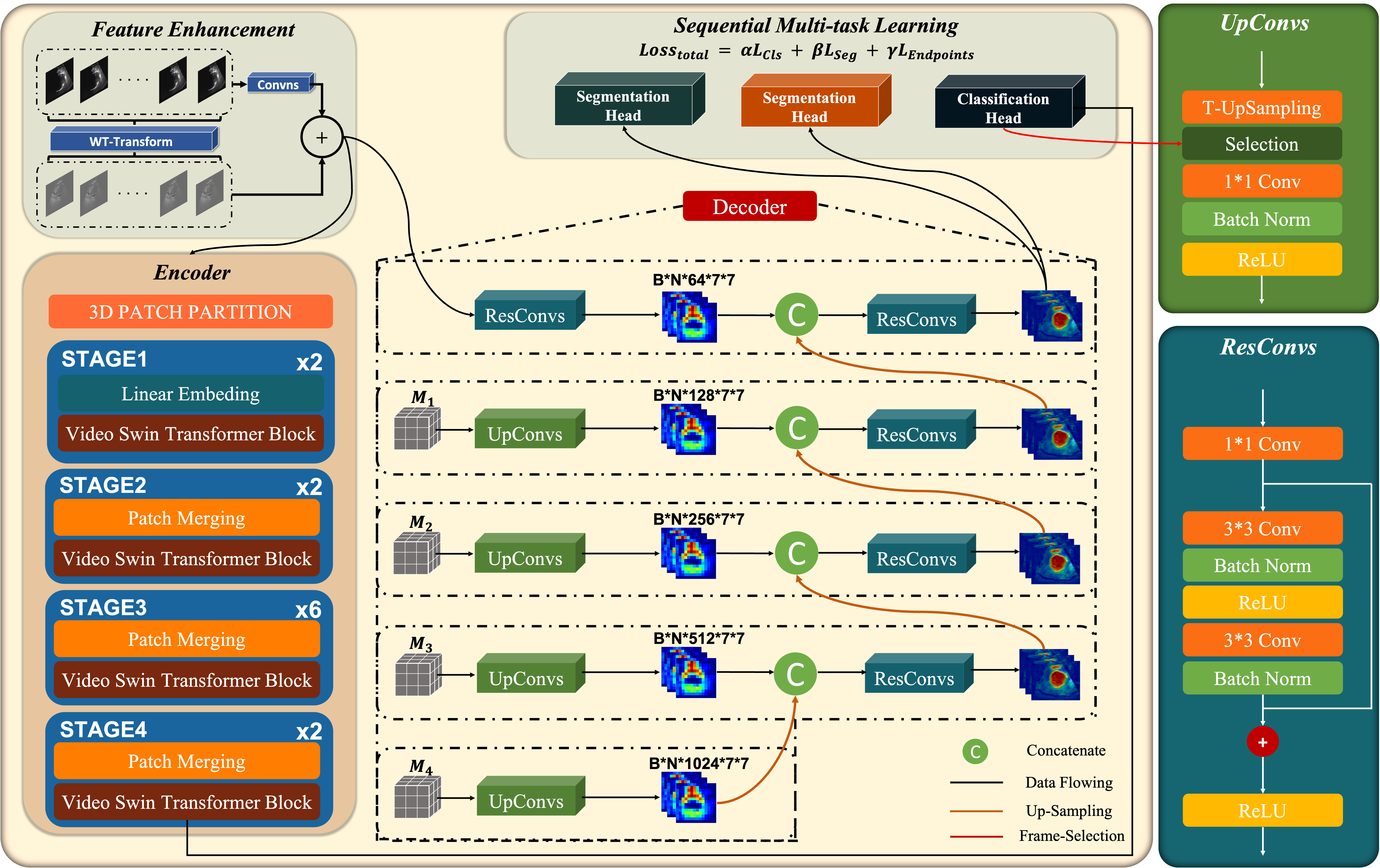}
  \caption{The proposed SSTN includes a feature enhancement block, a Video Swin Transformer as the encoder, and a decoder with ResConvs and UpConvs for optimized feature extraction. The Sequential Multi-task Learning framework uses a weighted loss function to balance classification, segmentation, and landmark detection tasks.}
  \label{fig:pipeline}
\end{figure*}
Ultrasound imaging provides real-time, non-invasive visualization of internal anatomy, proven to offer a higher degree of accuracy than digital palpation in monitoring fetal head descent. Multiple assessments are often required to track fetal head position, enabling timely diagnostics and interventions. According to the guideline from International Society of Ultrasound in Obstetrics and Gynecology (ISUOG), it highlights the Angle of Progression (AoP) and Head-Symphysis Distance (HSD) — derived from measurements of the fetal head and pubic symphysis — as key indicators for predicting successful instrumental delivery \cite{ghi2018isuog,youssef2013fetal,youssef2017automated}. Notably, AoP is the angle between the pubic bone and lowest edge of pubis tangentially drawn to run along the deepest part of the fetal head \cite{ghi2018isuog,yeo2009sonographic}. Specifically, there is an ultrasound probe that is placed on maternal suprapubic region to identify the fetal head. Landmarks to identify orientation such as fetal orbits or occiput are then noted. The fetal head midline features become difficult to determine transabdominally with fetal descent. A trans perineal approach is then recommended for precise determination of position. In this process, the pubic symphysis is used as a landmark for quantitive measurements. The process is illustrated in \cref{fig:aop&hsd}. Therefore, the structured approach can be formulated as a sequential process, including the identification of standardized ultrasound planes, the segmentation of anatomical structures, and the detection of key landmarks for precise measurement of AoP and HSD. The AoP and HSD are then derived based on the identified anatomical landmarks. The sequential process is essential for accurate and reliable measurements of AoP and HSD, as well as for the interpretation of the results. It enhances the robustness of the measurement process and improves the interpretability of the biometric assessment.

Numerous studies have addressed structured approaches in ultrasound image analysis within labor environments \cite{lu2022multitask,ou2024rtseg,chen2024psfhs,qiu2024psfhsp,bai2022framework,zhou2023segmentation, ramesh2024geometric}. However, these studies treat the tasks as independent and parallel processes. It may lead to suboptimal results and against to clinical practice, as the tasks are inherently interrelated and dependent on one another. Besides, the low signal-noise-rate characteristic in ultrasonic samples is not neglectable. To leverage the temporal information from the ultrasound video, LSTM networks can be adopted but are limited in capturing global features \cite{cheng2022xmem,Deng_2024_CVPR,10.1007/978-3-031-52388-5_9}.

To address the challenges in the structured approach for intrapartum ultrasound videos analysis, we propose the Sequential Spatial-Temporal Network (SSTN) for the automatic assessment of fetal head descent during labor. The sequential processing of SSTN starts with identifying the ultrasound planes, followed by segmenting the anatomical structures, and finally detecting key landmarks for precise measurement of AoP and HSD. To our best knowledge, SSTN is the first interpretable model specifically designed for the video of intrapartum ultrasound analysis. Besides, the SSTN is capable of capturing temporal, global, and local information from both the preceding and subsequent frames of the ultrasound video. Experiment results show that SSTN significantly surpasses existing models, reducing the mean absolute error by 18\% for AoP and 22\% for HSD. It demonstrates the effectiveness of SSTN in dealing with specific labor assessment tasks.

\section{Method}
\label{sec:method}
\subsection{SSTN}
\label{SSTN}
In this section, we formally introduce SSTN for the automatic assessment of fetal head descent during labor. The SSTN architecture comprises three primary components: a feature enhancement block, a Video Swin Transformer (VST) encoder, and a decoder with ResConvs and UpConvs for optimized feature extraction. The architecture is illustrated in \cref{fig:pipeline}.

\begin{table*}[ht]
  \centering
  \renewcommand{\arraystretch}{1.5} 
  \begin{tabular}{lcccclccclcc}
    \toprule
    \multirow{2}{*}{\textbf{Methods}}  & \multicolumn{4}{c}{\textbf{Classification (STAGE 1)}} &                    & \multicolumn{3}{c}{\textbf{Segmentation (STAGE 2)}} &                & \multicolumn{2}{c}{\textbf{Biometry (STAGE 3)}}                                                                                                                   \\ \cline{2-5} \cline{7-9} \cline{11-12}
                                       & ACC$\uparrow$                                         & F1-Score$\uparrow$ & AUC$\uparrow$                                       & MCC$\uparrow$  & \textbf{}                                       & DSC$\uparrow$  & HD$\downarrow$  & ASD$\downarrow$ &  & ${\Delta }$AoP$\downarrow$ & ${\Delta }$HSD$\downarrow$ \\ \cline{1-5} \cline{7-9} \cline{11-12}
    Baseline                           & 0.492                                                 & 0.450              & \textbf{0.792}                                      & 0.250          &                                                 & 0.844          & 48.182          & 10.861          &  & 11.351                     & 14.150                     \\
    Swin-UNet                          & \textbf{0.856}                                        & 0.149              & 0.712                                               & 0.533          &                                                 & 0.864          & 37.511          & 10.680          &  & 14.150                     & 14.081                     \\ \hline
    \textit{w/o} FE \textit{w/o} ST & --                                                    & --                 & --                                                  & --             &                                                 & --             & --              & --              &  & 13.601                     & 12.156                     \\
    \textit{w/} FE \textit{w/o} ST  & --                                                    & --                 & --                                                  & --             &                                                 & --             & --              & --              &  & 9.705                      & 9.896                      \\
    \textit{w/o} FE \textit{w/} ST  & 0.792                                                 & 0.801              & 0.562                                               & 0.780          &                                                 & 0.884          & 35.612          & 9.280           &  & 8.950                      & 9.681                      \\ \hline
    \textbf{SSTN}                      & 0.821                                                 & \textbf{0.817}     & 0.656                                               & \textbf{0.835} &                                                 & \textbf{0.901} & \textbf{33.037} & \textbf{8.263}  &  & \textbf{8.257}             & \textbf{9.489}             \\ \bottomrule
  \end{tabular}
  \caption{Experiment results of methods on sequential tasks in labor diagnoses, including classification, segmentation, and biometry stages. The unit for ${\Delta \text{AoP}}$ is $^\circ$ and the unit for ${\Delta \text{HSD}}$ is pixels.}
  \label{tab:exp}
\end{table*}

The video sequence $\mathcal{V}_{L}$, consisting of $L$ frames, through a preprocessing stage designed to enhance task-specific features, resulting in the enriched features $\mathcal{D}_{L}$. Specifically, we employ the WTConv2d layer \cite{finder2024wavelet} as the learnable feature enhancement block. It can mitigate the impact of noise on feature learning and extraction. Following the feature enhancement operation within the enhancement block, the enhanced feature maps, denoted as $\mathcal{V}_{\text{enhance}}$, are input into the VST to generate multi-level feature maps, $\mathcal{M}_{S}$, at each stage $\mathcal{S}$, prior to the patch merging process. $\mathcal{M}_{S}$ encapsulates rich spatial-temporal representations at various levels of abstraction. The final-stage feature map, $\mathcal{M}_{4}$, is then forwarded to the classification head, which comprises a 2D global average pooling layer followed by a fully connected layer, resulting in precise classification analysis. After completing the classification task, the decoder utilized for segmentation and landmark detection will share the same encoder, the VST. Except for the bottleneck feature map, $\mathcal{M}_{4}$, the remaining feature maps, $\mathcal{M}_{S}$, will provide multi-scale reconstruction information for segmentation and landmark detection tasks through a skip connection approach \cite{ronneberger2015u}, enhancing the model's robustness in these tasks. To further refine the feature maps, the ResConvs and UpConvs blocks are employed to preserve the spatial information according to the identified anatomical landmarks. During inference, UpConvs layers filter frames based on classification results, whereas during training, they filter frames based on only annotated frame positions. The filtered frames are concatenated with the upsampled outputs from the preceding layer, and the feature dimensions are adjusted through ResConvs blocks before additional up-sampling. Finally, the multi-scale feature maps are input into the landmark and segmentation heads to precisely measure the AoP and HSD.

\subsection{Supervision}
\label{ssec:loss}
The classification, segmentation, and landmark detection tasks are interrelated. The accurate measurement of the AoP and HSD by the model is contingent upon the extraction of Regions of Interest (ROIs) from the full pubic symphysis and fetal head within the standard ultrasound plane. Therefore, the classification task is prioritized over the segmentation and landmark detection tasks in the order of execution among these tasks. However, with reference to the computational indications of AoP and HSD provided by ISUOG is inherently influenced by the segmentation task's accuracy. Thus, it preserves the landmark information in previous detection to refine the segmentation. Specifically, the landmark detection task serves as a guiding function for the segmentation task. Therefore, to enable the model to achieve multitask learning, we perform a weighted summation of the loss functions associated with the three tasks, defined as follows:

\begin{equation}
  Loss_{total} = \alpha L_{Cls} + \beta L_{Seg} +  \gamma L_{Endpoints},
\end{equation}
where $\alpha$, $\beta$, and $\gamma$ are weighting coefficients for the classification, segmentation, and landmarks detection tasks, respectively. In this framework, binary cross-entropy (BCE) is employed as the loss function for the classification task. For the segmentation task, which requires precise delineation of the pubic symphysis and fetal head, a composite loss function comprising the Dice score (DSC), BCE, and Intersection over Union (IoU) is utilized to optimize both boundary accuracy and region overlap. The endpoint detection task leverages the mean squared error (MSE) as its loss function, aimed at minimizing the error in landmark detection. The specific formulation for segmentation loss is defined as follows:

\begin{equation}
  L_{Seg} = \eta L_{IoU} + \epsilon L_{Dice} + \xi L_{CE} ,
\end{equation}
where $\eta$, $\epsilon$, and $\xi$ are weighting coefficients for each respective loss function.

\section{EXPERIMENTS AND RESULTS}
\label{sec:exp}

\subsection{Dataset and Implementation}
The dataset sources from the Intrapartum Ultrasound Grand Challenge\footnote{\url{https://codalab.lisn.upsaclay.fr/competitions/18413}}, comprising 435 sample patients, with one video for one sample, totaling 54,002 frames. Only 2,700 frames are annotated for segmentation tasks. Initially, each frame is resized to 224$\times$224. Prior to inputting the data into the SSTN, the frames are processed into video clips $\mathcal{V}_{L}$, with $L$ set to 4 to balance computational efficiency while capturing dynamic noise features across consecutive frames. The learning rate is set to $10^{-4}$. $\alpha$, $\beta$, and $\gamma$ are set to 1, 10, and 10, respectively. SSTN is implemented via PyTorch.

\subsection{Baseline Results}
\label{ssec:task1}
We conduct a rigorous comparative experiment of complex labor diagnoses to comprehensively validate the enhanced robustness and stability of the proposed SSTN. Our method was benchmarked against widely used techniques across different medical applications, focusing on classification, segmentation, and landmark detection tasks. For the classification task, we implement ResNet-18 \cite{he2016deep}, the Swin Transformer \cite{liu2021Swin}, and a baseline (UNet-Encoder) \cite{cao2022swin}, evaluating performance through accuracy (ACC), F1-score, Matthews correlation coefficient (MCC), and area under the curve (AUC). In the segmentation task, our model was compared with the baseline (UNet) and Swin-UNet, using Intersection over Union (IoU), Dice Similarity Coefficient (DSC), Hausdorff Distance (HD), and Average Surface Distance (ASD) to assess segmentation accuracy. For the biometry task, both baseline and Swin-UNet obtained predictions directly from the ISUOG-recommended calculations for AoP and HSD. At the same time, our method predicted these parameters by identifying key anatomical landmarks in each standard plane.

\begin{figure}[htp]
  \centering
  \includegraphics[width=0.48\textwidth]{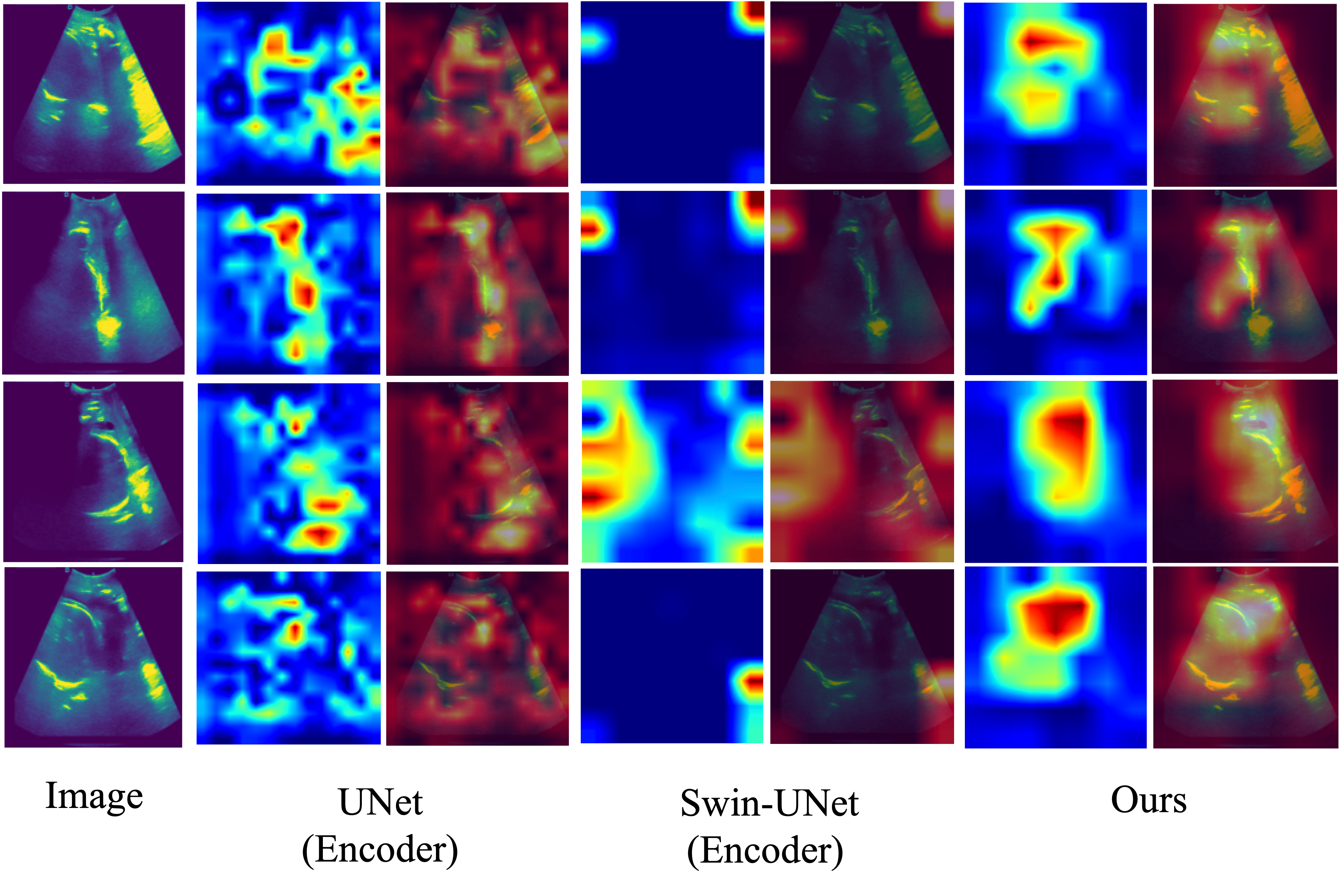}
  \caption{The features extracted by various models during the classification task can be visualized using Gradient-weighted Class Activation Mapping (Grad-CAM).}
  \label{fig:gradcam}
\end{figure}

\begin{figure}[htb]
  \centering
  \includegraphics[width=0.49\textwidth]{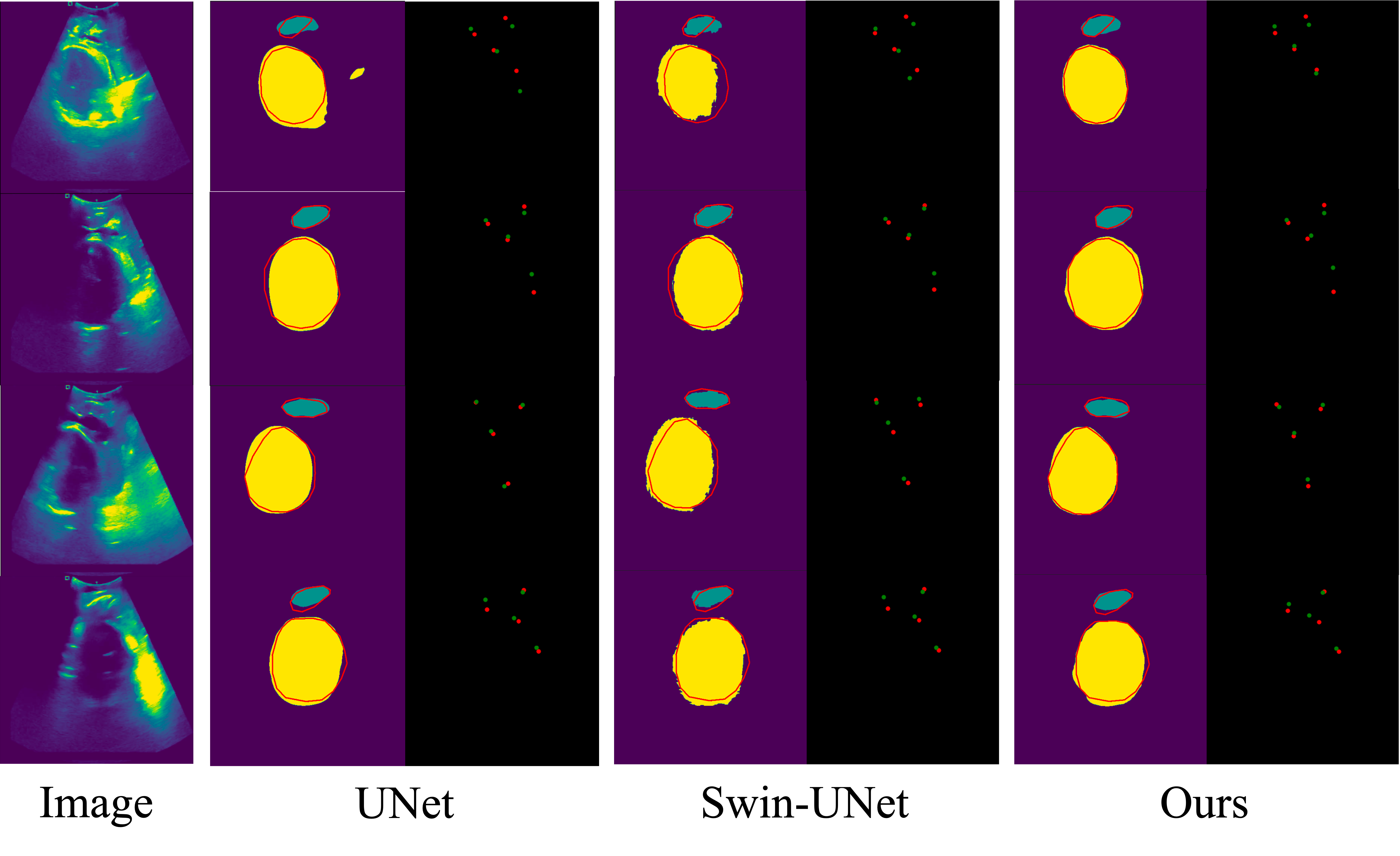}
  \caption{Segmentation (left) and keypoint (right) prediction results for various methods. In segmentation, the blue region marks the pubic symphysis, yellow indicates the fetal head, and the red curve represents the ground truth. For keypoint predictions, red points denote the ground truth, while green points show predicted keypoints.}
  \label{fig:seg&kp}
\end{figure}

\Cref{tab:exp} presents the performance of the SSTN model in comparison to other methods, achieving a classification accuracy of 82.0\% and an F1 score of 81.7\%. These results highlight the model's capability in effectively distinguishing clinically relevant features in ultrasound images (see \cref{fig:gradcam}). Furthermore, the results in \cref{tab:exp} demonstrate that the model achieves a DSC of 0.910 in the segmentation task, representing a 7.82\% improvement over the baseline. Notably, the error rates for ${\Delta \text{AoP}}$ and ${\Delta \text{HSD}}$ are reduced by 18\% and 22\%, respectively. This suggests that robust classification ensures the extraction of features that are both meaningful and clinically relevant, thereby enhancing the model's interpretability and stability (see \cref{fig:seg&kp}). As a result, the model's improved performance in segmentation and biometry tasks is directly influenced by the quality of feature extraction. Additionally, the joint training of classification, segmentation, and biometry tasks facilitates the simultaneous learning of critical features, further enhancing the model's overall effectiveness in noisy and complex ultrasound environments.

\subsection{Ablation Analysis}
\label{ssec:ablation}
To further validate the effectiveness of our method, we conducted an ablation study to evaluate the contribution of each module within our model, as shown in the lower part of Table 1. The study investigates three variants of SSTN, each representing different combinations. The first variants without sequential task (ST) processing and the feature enhance block (FE) is not able to evaluate the performance on stage 1 and stage 2, but still can estimate the AoP and HSD. The second variant with the FE block but without ST, and the third only with ST. The results highlight the significant advantages brought by FE and ST, exhibiting superior performance and enhanced robustness of the proposed model.

\section{Conclusion}
\label{sec:majhead}
In this study, we introduce the Sequential Spatial-Temporal Network (SSTN), the first interpretable model for the analysis of intrapartum ultrasound video. The sequential process of SSTN preserves the previous stage information, guiding the next stage task. It maintains the results that are interpretable and transparent by explicitly following the practical guideline. Multitask supervision also provides sufficient constraints on prototypes of AoP and HSP estimation. The results of extensive experiments show our approach superior efficiency in dealing with specific labor assessment tasks. The ablation study further demonstrates the advantages of sequential processing and characteristic feature enhancement. Because of the interpretable structure, SSTN is beneficial to practical clinical scenarios. We wish that SSTN can be further studied and benefit future research work for better labor assessment.

\vfill
\pagebreak

\section{COMPLIANCE WITH ETHICAL STANDARDS}
This research study was conducted retrospectively using human subject data made available in open access by \cite{CHEN2024123096,bai2022framework,LU2022107904,lu2022multitask}. Ethical approval was not required as confirmed by the license
attached with the open access data.

{\setstretch{0.9}
\bibliographystyle{IEEEbib}
\bibliography{refs}

\begin{thebibliography}{10}

\bibitem{ghi2018isuog}
T~Ghi, T~Eggeb{\o}, C~Lees, et~al.,
\newblock ``Isuog practice guidelines: intrapartum ultrasound,''
\newblock {\em Ultrasound in Obstetrics \& Gynecology}, vol. 52, no. 1, pp. 128--139, 2018.

\bibitem{youssef2013fetal}
A~Youssef, E~Maroni, A~Ragusa, et~al.,
\newblock ``Fetal head--symphysis distance: a simple and reliable ultrasound index of fetal head station in labor,''
\newblock {\em Ultrasound in Obstetrics \& Gynecology}, vol. 41, no. 4, pp. 419--424, 2013.

\bibitem{youssef2017automated}
Aly Youssef, Ginevra Salsi, Elisa Montaguti, et~al.,
\newblock ``Automated measurement of the angle of progression in labor: a feasibility and reliability study,''
\newblock {\em Fetal Diagnosis and Therapy}, vol. 41, no. 4, pp. 293--299, 2017.

\bibitem{yeo2009sonographic}
Lami Yeo and Roberto Romero,
\newblock ``Sonographic evaluation in the second stage of labor to improve the assessment of labor progress and its outcome,''
\newblock {\em Ultrasound in obstetrics \& gynecology: the official journal of the International Society of Ultrasound in Obstetrics and Gynecology}, vol. 33, no. 3, pp. 253, 2009.

\bibitem{lu2022multitask}
Yaosheng Lu, Dengjiang Zhi, Minghong Zhou, et~al.,
\newblock ``Multitask deep neural network for the fully automatic measurement of the angle of progression,''
\newblock {\em Computational and Mathematical Methods in Medicine}, vol. 2022, no. 1, pp. 5192338, 2022.

\bibitem{ou2024rtseg}
Zhanhong Ou, Jieyun Bai, Zhide Chen, et~al.,
\newblock ``Rtseg-net: a lightweight network for real-time segmentation of fetal head and pubic symphysis from intrapartum ultrasound images,''
\newblock {\em Computers in Biology and Medicine}, vol. 175, pp. 108501, 2024.

\bibitem{chen2024psfhs}
Gaowen Chen, Jieyun Bai, Zhanhong Ou, et~al.,
\newblock ``Psfhs: intrapartum ultrasound image dataset for ai-based segmentation of pubic symphysis and fetal head,''
\newblock {\em Scientific Data}, vol. 11, no. 1, pp. 436, 2024.

\bibitem{qiu2024psfhsp}
Ruiyu Qiu, Mengqiang Zhou, Jieyun Bai, et~al.,
\newblock ``Psfhsp-net: an efficient lightweight network for identifying pubic symphysis-fetal head standard plane from intrapartum ultrasound images,''
\newblock {\em Medical \& Biological Engineering \& Computing}, pp. 1--12, 2024.

\bibitem{bai2022framework}
Jieyun Bai, Zhanhang Sun, Sheng Yu, et~al.,
\newblock ``A framework for computing angle of progression from transperineal ultrasound images for evaluating fetal head descent using a novel double branch network,''
\newblock {\em Frontiers in Physiology}, vol. 13, pp. 940150, 2022.

\bibitem{zhou2023segmentation}
Mengqiang Zhou, Chuan Wang, Yaosheng Lu, et~al.,
\newblock ``The segmentation effect of style transfer on fetal head ultrasound image: a study of multi-source data,''
\newblock {\em Medical \& Biological Engineering \& Computing}, vol. 61, no. 5, pp. 1017--1031, 2023.

\bibitem{ramesh2024geometric}
Jayroop Ramesh, Nicola Dinsdale, Pak-Hei Yeung, et~al.,
\newblock ``Geometric transformation uncertainty for improving 3d fetal brain pose prediction from freehand 2d ultrasound videos,''
\newblock in {\em Proceedings of the International Conference on MICCAI}. Springer, 2024, pp. 419--429.

\bibitem{cheng2022xmem}
Ho~Kei Cheng and Alexander~G Schwing,
\newblock ``Xmem: Long-term video object segmentation with an atkinson-shiffrin memory model,''
\newblock in {\em Proceedings of the ECCV}. Springer, 2022, pp. 640--658.

\bibitem{Deng_2024_CVPR}
Xiaolong Deng, Huisi Wu, Runhao Zeng, et~al.,
\newblock ``Memsam: Taming segment anything model for echocardiography video segmentation,''
\newblock in {\em Proceedings of the IEEE/CVF CVPR}, June 2024, pp. 9622--9631.

\bibitem{10.1007/978-3-031-52388-5_9}
Hamza Hadri, Abderahhim Fail, and Mohamed Sadik,
\newblock ``Ultrasound beamforming: Investigating time series with sequence to sequence approach in deep learning,''
\newblock in {\em International Conference on Advanced Intelligent Systems for Sustainable Development}, Cham, 2024, pp. 88--97, Springer Nature Switzerland.

\bibitem{finder2024wavelet}
Shahaf~E Finder, Roy Amoyal, Eran Treister, et~al.,
\newblock ``Wavelet convolutions for large receptive fields,''
\newblock in {\em Proceedings of the ECCV}, 2024.

\bibitem{ronneberger2015u}
Olaf Ronneberger, Philipp Fischer, and Thomas Brox,
\newblock ``U-net: Convolutional networks for biomedical image segmentation,''
\newblock in {\em Proceedings of MICCAI}. Springer, 2015, pp. 234--241.

\bibitem{he2016deep}
Kaiming He, Xiangyu Zhang, Shaoqing Ren, et~al.,
\newblock ``Deep residual learning for image recognition,''
\newblock in {\em Proceedings of the IEEE/CVF CVPR}. IEEE, 2016, pp. 770--778.

\bibitem{liu2021Swin}
Ze~Liu, Yutong Lin, Yue Cao, et~al.,
\newblock ``Swin transformer: Hierarchical vision transformer using shifted windows,''
\newblock in {\em Proceedings of the IEEE/CVF ICCV}, 2021.

\bibitem{cao2022swin}
Hu~Cao, Yueyue Wang, Joy Chen, et~al.,
\newblock ``Swin-unet: Unet-like pure transformer for medical image segmentation,''
\newblock in {\em Computer Vision -- ECCV 2022 Workshops}, Cham, 2023, pp. 205--218, Springer Nature Switzerland.

\bibitem{CHEN2024123096}
Zhensen Chen, Zhanhong Ou, Yaosheng Lu, et~al.,
\newblock ``Direction-guided and multi-scale feature screening for fetal head–pubic symphysis segmentation and angle of progression calculation,''
\newblock {\em Expert Systems with Applications}, vol. 245, pp. 123096, 2024.

\bibitem{LU2022107904}
Yaosheng Lu, Mengqiang Zhou, Dengjiang Zhi, et~al.,
\newblock ``The jnu-ifm dataset for segmenting pubic symphysis-fetal head,''
\newblock {\em Data in Brief}, vol. 41, pp. 107904, 2022.

\end{thebibliography}
}
\end{document}